\documentclass[a4paper,11pt]{article}
\usepackage{pos}
\usepackage{tikz}
\usepackage{calc}
\usetikzlibrary{shapes,arrows}
\usetikzlibrary{fit,calc}
\usetikzlibrary{positioning}

\newcommand{\pp}{\ensuremath{\text{p\kern-0.05em p}}}

\newcommand{\PbPb}{\ensuremath{\mbox{Pb--Pb}}}

\newcommand{\sqrts}{\ensuremath{\sqrt{s_{\text{NN}}}}}

\newcommand{\figRef}[1]{Fig.~\ref{#1}}
\newcommand{\figureRef}[1]{Figure~\ref{#1}}

\newcommand{\GeVc}{\ensuremath{\text{GeV}\kern-0.05em/\kern-0.02em c}}

\newcommand{\pT}{\ensuremath{p_{\text{T}}}}

\newcommand{\pTJet}{\ensuremath{p_{\text{T,jet}}}}

\newcommand{\pTJetCh}{\ensuremath{p_{\text{T,jet}}^{\text{ch}}}}

\newcommand{\zg}{\ensuremath{z_{\text{g}}}}
\newcommand{\zcut}{\ensuremath{z_{\text{cut}}}}
\newcommand{\Rg}{\ensuremath{R_{\text{g}}}}

\newcommand{\kT}{\ensuremath{k_{\text{T}}}}
\newcommand{\kTg}{\ensuremath{k_{\text{T,g}}}}
\newcommand{\deltaR}{\ensuremath{\Delta R}}

\title{Exploring jet interactions in the quark-gluon plasma using jet substructure measurements in Pb-Pb collisions with ALICE}
\ShortTitle{Exploring jet interactions via jet substructure with ALICE}

\manuallySeparateAuthors{}
\author*{Raymond Ehlers}
\author{on behalf of the ALICE Collaboration}

\affiliation{Lawrence Berkeley National Laboratory/UC Berkeley}

\emailAdd{raymond.ehlers@cern.ch}

\abstract{
Jets are generated in hard interactions in high-energy nuclear collisions.
Jets propagate through the quark-gluon plasma (QGP) as the jet shower evolves;
their interaction with the QGP, known as jet quenching, generates observable phenomena that provide incisive probes of the structure and dynamics of the QGP.
For instance, medium-induced modification of jet substructure probes color coherence, and may be sensitive to differences in quark and gluon energy loss due to their different Casimir factors.
Jet grooming can be used to focus on specific regions of phase space, isolating medium-induced effects on hard splittings in the jet shower.
ALICE is well suited for such substructure measurements due to its precise charged-particle tracking, which enables high-efficiency measurements of narrow splittings in jets down to low transverse momentum.
In these proceedings several recent jet substructure measurements in Pb--Pb collisions at $\sqrt{s_{\mathrm{NN}}} = 5.02$ TeV are reported, for both ungroomed jets and jets that have been groomed using the Soft Drop and Dynamical Grooming algorithms.
Measurements of the groomed jet radius, $\theta_g \equiv R_g/R$; the groomed jet momentum fraction, $z_g$; and the groomed relative transverse momentum, $k_{\mathrm{T,g}}$ are reported.
These measurements show direct evidence of modification of the angular structure of jets in the QGP, and provide new constraints on the search for large-angle scattering of jets off of quasi-particles by interaction with the QGP.
New measurements of sub-jet fragmentation, generalized jet angularities, and jet-axis differences, which provide insight into the angular and momentum structure of modified jets are also presented.
Comparisons to model calculations are discussed.
}

\FullConference{%
  41st International Conference on High Energy physics - ICHEP2022\\
  6-13 July, 2022\\
  Bologna, Italy
}

\begin{document}
\maketitle

\hypertarget{introduction}{%
\section{Introduction}\label{introduction}}

High momentum-transfer processes in high energy nuclear collisions
generate collimated sprays of particles known as jets. As jets propagate
through the quark-gluon plasma (QGP) formed in these collisions, their
interactions encode properties of the medium via modification of jet properties and structure, in a
process known as jet quenching. The measurement of jet substructure observables, and their comparison to theoretical calculations,
can isolate and quantify these
modifications, providing insight into QGP properties.

The ALICE experiment~\cite{Abelev:2014ffa} is well-suited for such substructure
measurements. ALICE is able to measure small jet splitting angles with high precision, due to precise charged-particle tracking down to low
particle transverse momentum (\(\pT{}\)) using the Inner Tracking System and
Time Projection Chamber. ALICE also measures jets to the lowest
\(\pTJet{}\) at the LHC, where modifications due to the medium are
largest relative to the momentum scale of the probe. 
These capabilities enable a broad jet substructure program.

Jet are reconstructed using the anti-\(\kT{}\) algorithm implemented in
FastJet \cite{Cacciari:2011ma}. All results presented here are based on
charged-particles jets, which utilize charged particles
for jet reconstruction. Jet finding in \(\pp{}\) and \(\PbPb{}\) collisions
at \(\sqrts{}\) = 5.02 TeV is performed using jet resolution
parameters \(R\) = 0.2 and 0.4. The data were recorded during the 2017 and 2018 LHC runs. Jet
candidates selected for further analysis are required to be contained
fully within the fiducial acceptance, with background contributions in
\(\PbPb{}\) collisions subtracted via constituent subtraction
\cite{Berta:2019aa}.

After jet selection, jet substructure is measured using both groomed
and ungroomed approaches. Groomed substructure removes soft, wide-angle
emissions to isolate the hardest splittings, while
ungroomed substructure probes the entire splitting phase space,
including soft emissions. In the groomed approach, each jet is first
reclustered using the Cambridge-Aachen (C/A) algorithm. This declustered
splitting tree is then used as input for a grooming algorithm, which
searches through the tree and selects splittings according to specific criteria. For the results presented in these proceedings, groomed observables utilize
either 1) the Soft Drop (SD) grooming algorithm with a minimum shared
momentum fraction requirement, \(\zcut{}\) \cite{Larkoski:2014wba}, or
2) the Dynamical Grooming (DyG) algorithm with dynamic hardness measure
\(a\), which can be used to focus on, e.g.,~the largest \(\kT{}\)
(\(a = 1.0\)) \cite{Mehtar-Tani:2019rrk}. The observable is then
calculated using the properties of the selected splitting. In the
ungroomed approach, the jet constituents or declustered splitting tree
are directly analysed to calculate the observable of interest. In both
cases, observables are corrected for detector effects and background
fluctuations with two-dimensional unfolding in \(\pTJet{}\) and the
substructure observable using Bayesian iterative unfolding implemented in
RooUnfold \cite{Adye:1349242}.

\hypertarget{summary-substructure-observables}{%
\section{Summary substructure
observables}\label{summary-substructure-observables}}

Substructure observables which characterize the substructure of the entire jet via a single value are referred to as ``summary substructure'' observables. One such example is the angle between jet axes determined using
different jet recombination and grooming algorithms \cite{Cal:2019gxa}.
By using the entire jet, this observable is sensitive to the distribution
of soft radiation at large angles. To construct this observable,
charged-particle jets are reconstructed as usual using \(E\)
recombination scheme, which defines the standard jet axis. Each jet is
then reclustered using the C/A algorithm, utilizing either SD with
\(\zcut{} >\) 0.1 or 0.2 to define the SD axis, or alternatively, the
Winner-Take-All (WTA) \(\pT{}\) recombination scheme. The observable,
\(\Delta R_{\text{axis}} = \sqrt{(y_1 - y_2)^{2} + (\varphi_1 - \varphi_2)^2}\),
is then defined as the angular difference between the directions of two axes.

\begin{figure}[t]
    \centering
    \includegraphics[width=0.48\textwidth]{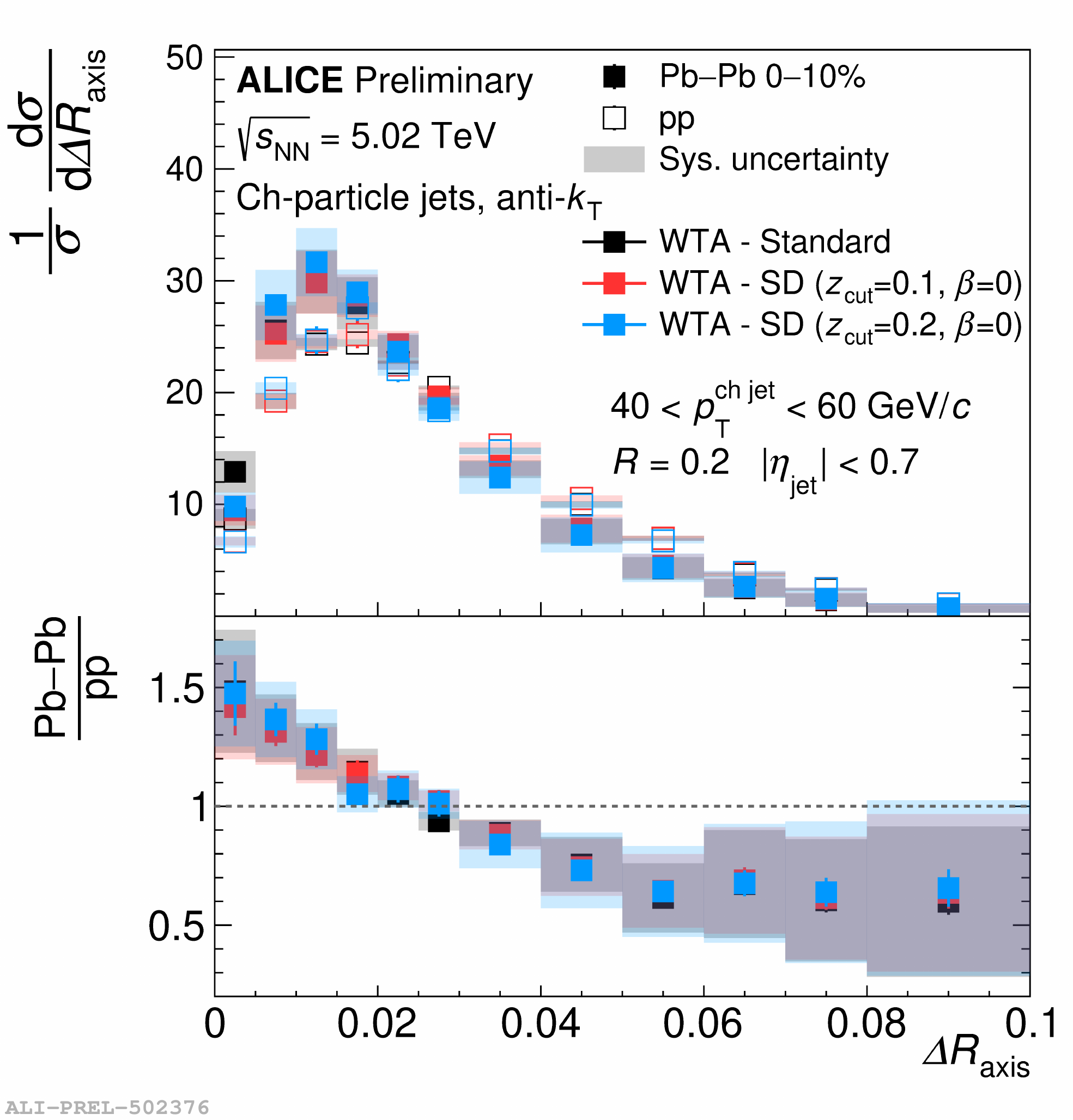}
    \includegraphics[width=0.48\textwidth]{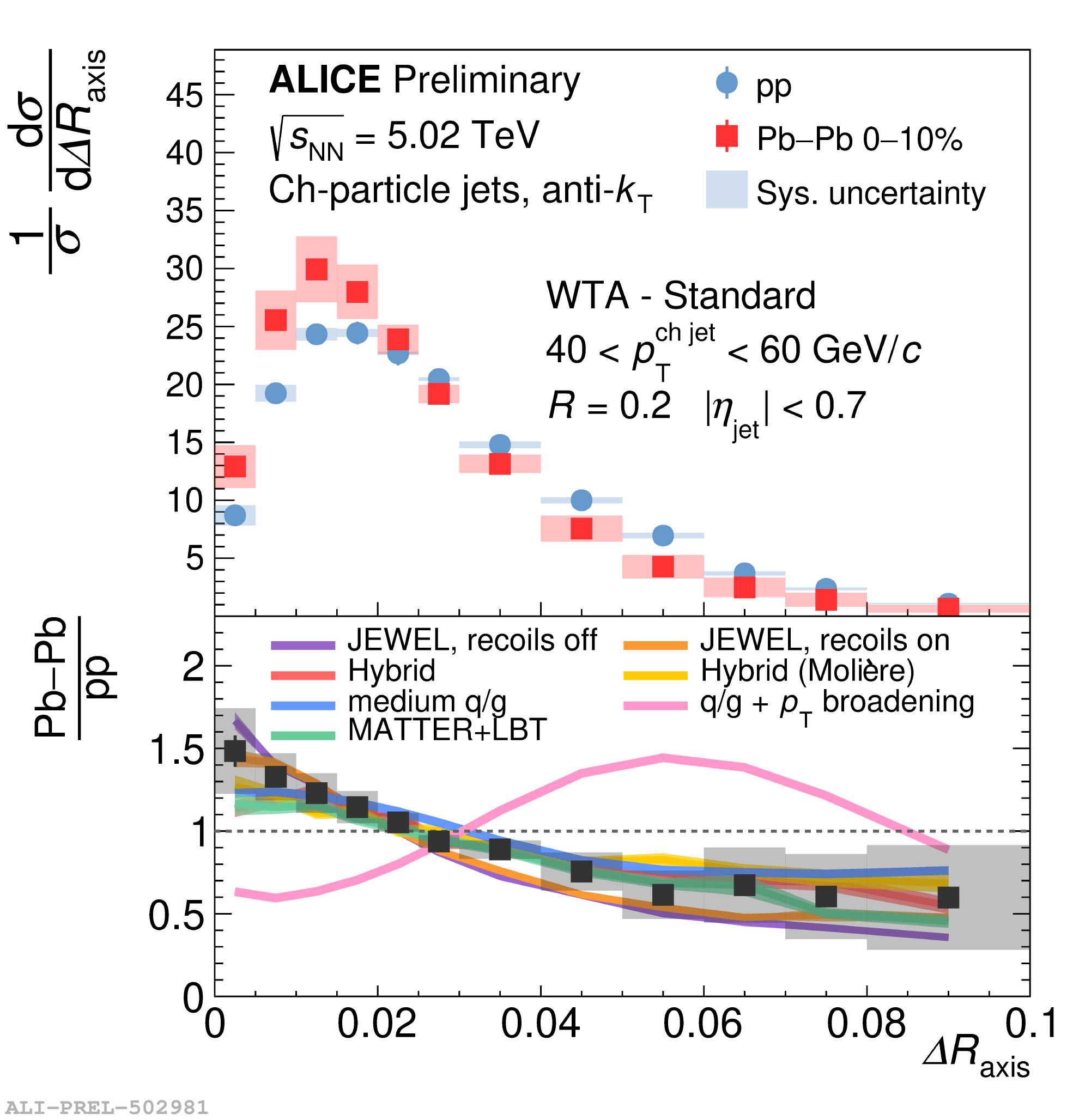}
    \caption{Jet axis difference for various axis definitions (left), compared to models for WTA vs standard jet axes (right), for charged-particle jets with $R = 0.2$ in $40 < \pTJetCh{} < 60$ \GeVc{} in \pp{} and 0--10\% central \PbPb{} collisions at \(\sqrts{}\) = 5.02 TeV.}
    \label{fig:jetAxis}
\end{figure}

Results are shown in \figRef{fig:jetAxis} for charged-particles jets with \(R\) = 0.2 in \(40 < \pTJetCh{} < 60\) \GeVc{}. In
all jet axis combinations (\figRef{fig:jetAxis}, left), the difference
between WTA and the other axis definitions is narrower in central (0--10\%)
\(\PbPb{}\) than in \(\pp{}\) collisions. The modification is qualitatively
described by a wide variety of models (\figRef{fig:jetAxis}, right)
despite those models utilizing different physics mechanisms to reproduce
the data, such as color coherence or changing the quark and gluon
fractions in the medium.

\begin{figure}[t]
    \centering
    \includegraphics[width=0.48\textwidth]{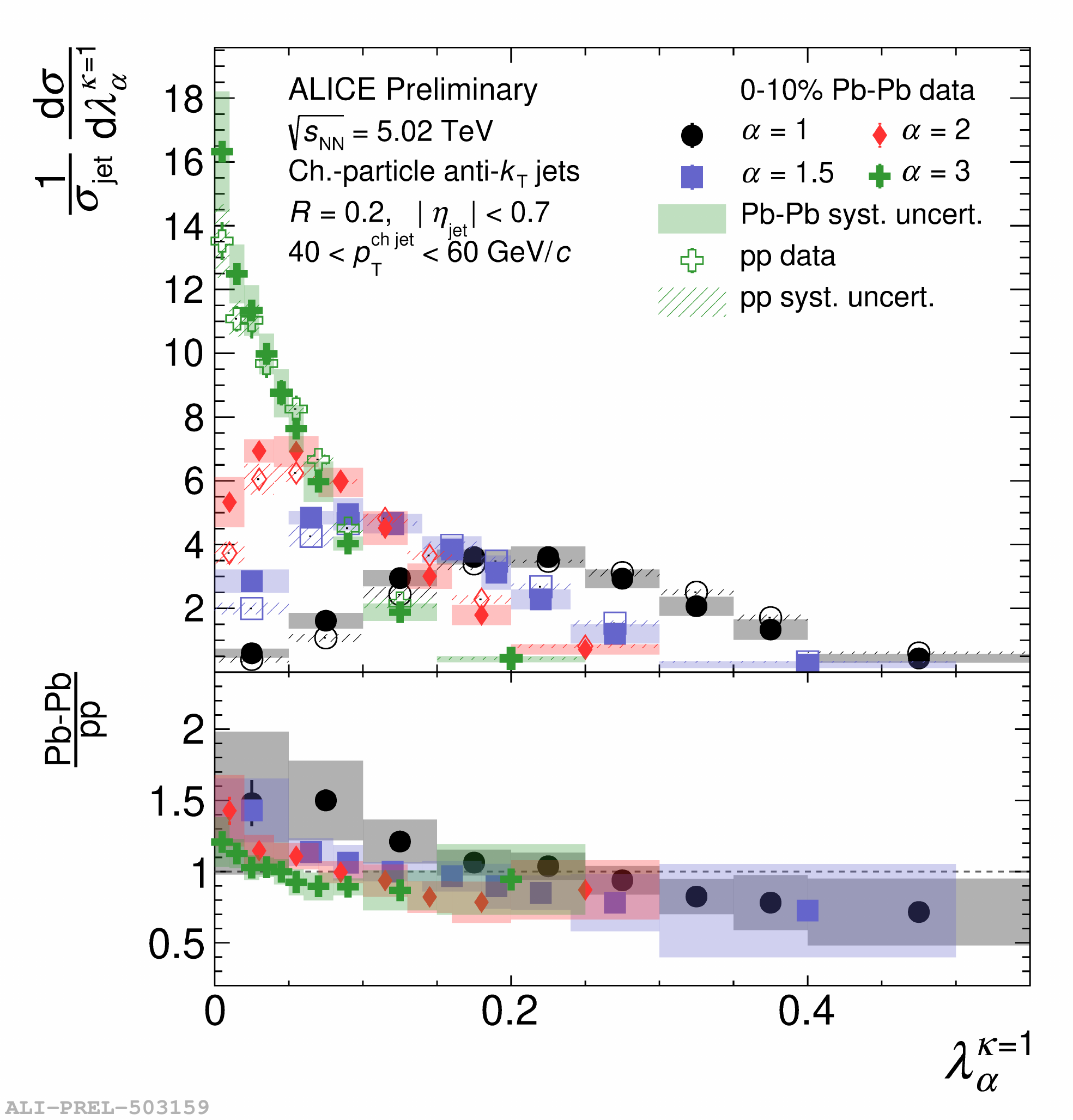}
    \includegraphics[width=0.48\textwidth]{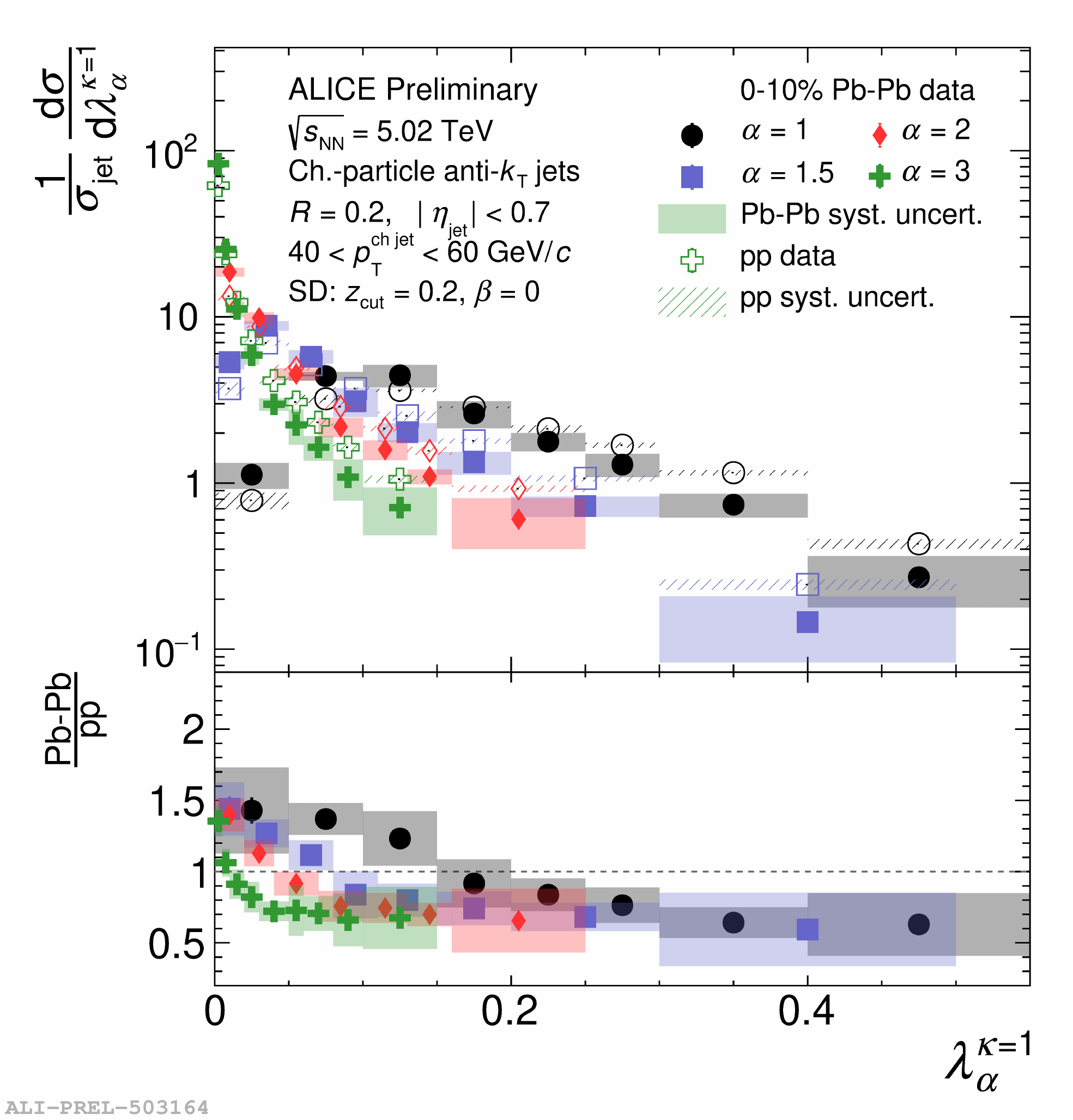}
    \caption{Generalized angularities for charged-particle jets with $R = 0.2$ in $40 < \pTJetCh{} < 60$ \GeVc{}, measured in $\pp{}$ and 0--10\% central $\PbPb{}$ collisions at \(\sqrts{}\) = 5.02 TeV. Ungroomed angularities are shown on the left, while groomed angularities are shown on the right.}
    \label{fig:generalizedAngularities}
\end{figure}

Jet substructure can also be characterized by generalized angularities,
which summarize the substructure via a \(\pT{}\)-weighed sum of the
angular distribution of the jet constituents. These angularities are
defined as
\(\lambda_{\alpha}^{\kappa} = \sum_{i} z_{i}^{\kappa} \theta_{i}^{\alpha}\),
where \(\alpha\) and \(\kappa\) are continuous parameters
\cite{Larkoski:2014pca}. The ALICE measurements are focused on
\(\kappa=1\) and \(\beta > 0\), which ensures infrared and collinear safety. For the
groomed case, which reduces intra-jet broadening and recoil effects, the
constituents of the subjet pair that passes the SD
condition of \(\zcut{} >\) 0.2 are used to calculate the
angularities. \figureRef{fig:generalizedAngularities} shows the ratio of
angularities with \(\alpha =\) 1, 1.5, 2, and 3 for \(R = 0.2\) jets with
\(40 < \pTJetCh{} < 60\) \GeVc{}, in central (0--10\%)
\(\PbPb{}\) and \(\pp{}\) collisions . Ungroomed angularities are shown on
the left, while groomed angularities are shown on the right. Both exhibit a narrowing
trend in \(\PbPb{}\) collisions, with smaller uncertainties and clearer narrowing
in the groomed case. Many models quantitatively describe the data (not
shown), with similar performance as for the jet axis difference.

\hypertarget{subjet-observables}{%
\section{Subjet observables}\label{subjet-observables}}

Another approach to studying jet substructure is to characterize the
properties of subjets found via declustering with the C/A algorithm. As
one example, ALICE has characterized the groomed momentum splitting
fraction \(\zg{}\) and jet radius \(\Rg{}\) for jets groomed with SD
\(\zcut{} > 0.2\) \cite{ALargeIonColliderExperiment:2021mqf}. This study
found no modification of \(\zg{}\), while \(\Rg{}\) was found to have an
enhancement of narrow splittings and a suppression of wide splittings,
corresponding to a promotion of narrow subjets or a filtering of wide
subjets. This narrowing is consistent with the jet axis difference and
generalized angularities.

Substructure can also be used to search for point-like (Moliere)
scattering centers in the medium~\cite{DEramo:2013aa,DEramo:2018eoy}.
Such scattering centers have been predicted to enhance the distribution of high relative transverse momentum emissions, \(\kT{}\), in \(\PbPb{}\) compared to
\(\pp{}\) collisions. To search for such an effect, we decluster each jet
candidate and utilize grooming algorithms to identify hard splittings,
including SD with \(\zcut{} > 0.2\), as well DyG
with \(a\) = 0.5, 1.0, and 2.0. For each identified splitting, the
observable is calculated as \(\kTg{} = p_{\text{T,2}} \sin{\deltaR}\),
where \(p_{\text{T,2}}\) is the \(\pT{}\) of the softer subjet and
\(\deltaR{}\) is the angle between the subjets.

\begin{figure}[t]
    \centering
    \includegraphics[width=0.48\textwidth]{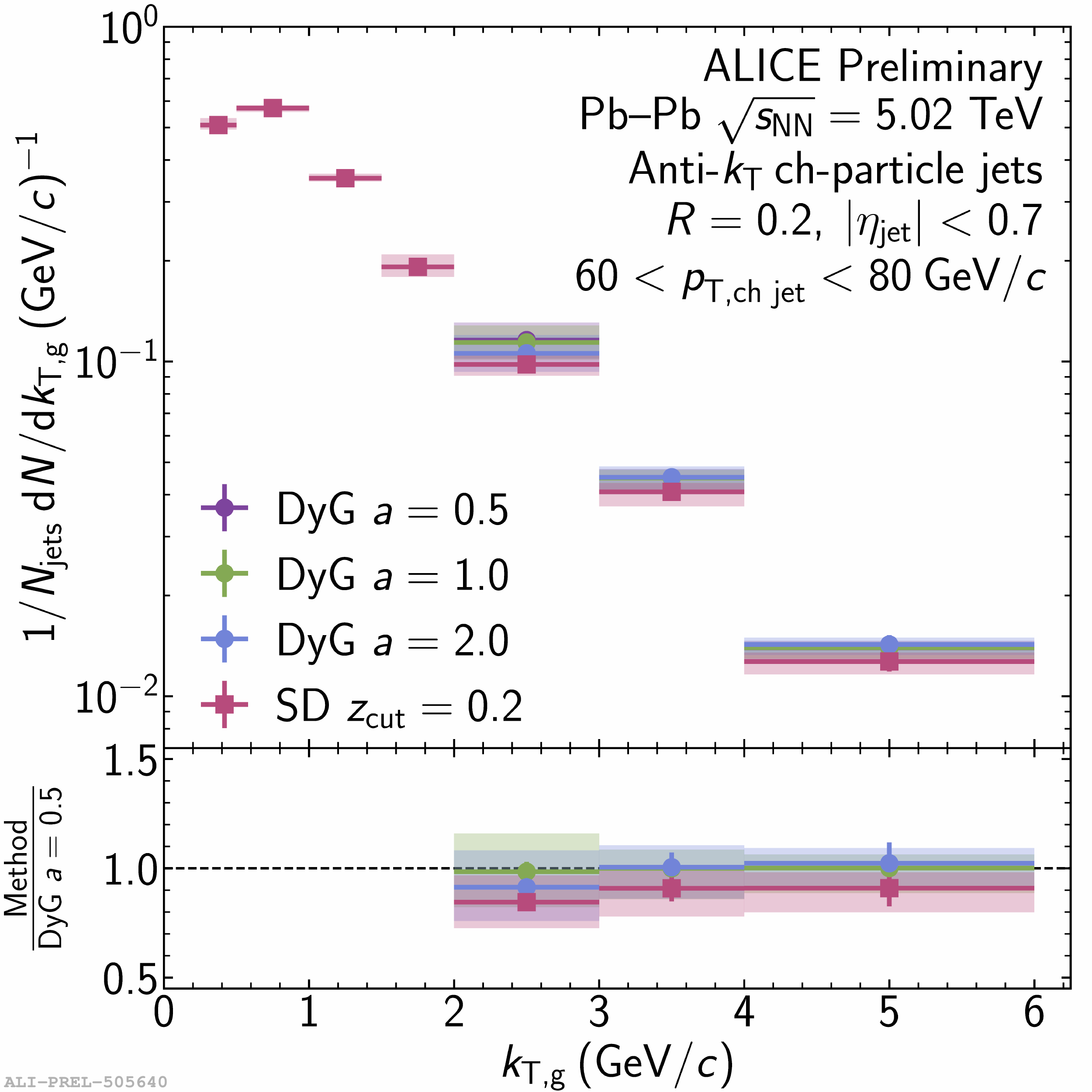}
    \includegraphics[width=0.48\textwidth]{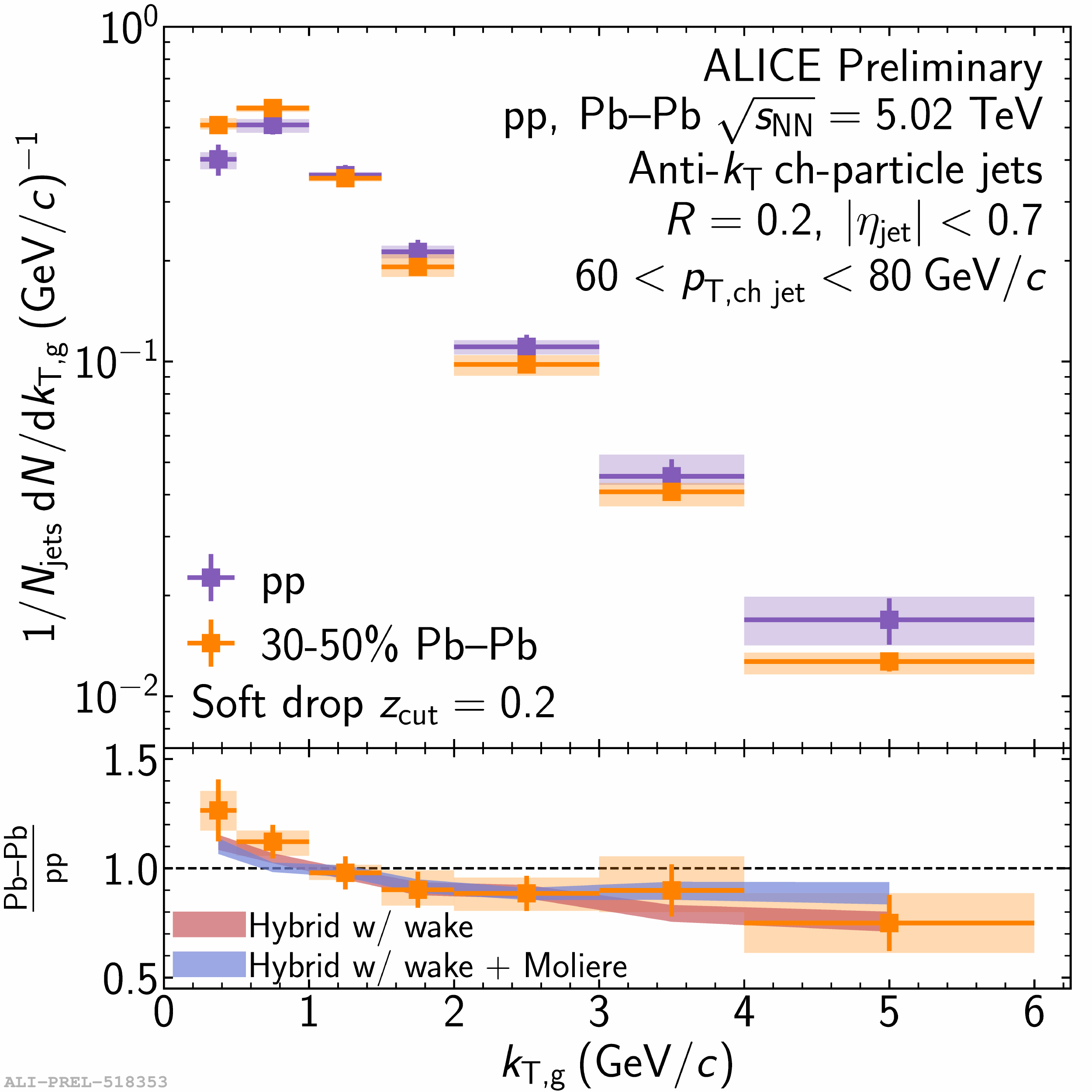}
    \caption{A comparison of hardest $\kTg{}$ splittings for a variety of grooming methods measured in 30--50\% semi-central $\PbPb{}$ collisions at \(\sqrts{}\) = 5.02 TeV is shown on the left, while a comparison of $\kTg{}$ measured in $\pp{}$ and $\PbPb{}$ collisions for Soft Drop with $\zcut{} > 0.2$ is shown on the right. Both panels show $R = 0.2$ charged-particle jets measured in $60 < \pTJetCh{} < 80$ \GeVc{}.}
    \label{fig:hardestKt}
\end{figure}

The measurement of \(\kTg{}\) using various grooming methods in 30--50\%
semi-central \PbPb{} collisions for \(R = 0.2\) charged-particle jets
measured in \(60 < \pTJetCh{} < 80\) \GeVc{} is shown on the left of
Fig. \ref{fig:hardestKt}. This is the first measurement using DyG in heavy-ion collisions, enabled by removing the soft \(\kT{}\)
contribution via the requirement of a minimum measured \(\kT{} > 1\)
\GeVc{}. All of the methods are consistent within uncertainties,
suggesting that they are selecting the same set of splittings at high
\(\kTg{}\). The right side of Fig. \ref{fig:hardestKt} shows a
comparison of the spectra measured in \(\pp{}\) and 30--50\% semi-central \(\PbPb{}\)
collisions for the same jet selections using SD with \(\zcut{} > 0.2\).
The \(\zcut{}\) in the SD grooming allows for the removal of the minimum
\(\kT{}\) requirement, which enables measuring over a larger range in
\(\kTg{}\). The apparent modification is consistent with the narrowing
observed in \(\Rg{}\), meaning that any possible point-like scattering
effects are convolved with this narrowing. Based on the comparison with
the Hybrid model \cite{DEramo:2018eoy}, this measurement does not yet have sufficient precision to be sensitive to the effects of Moliere scattering.

\begin{figure}[t]
    \centering
    \includegraphics[width=0.48\textwidth]{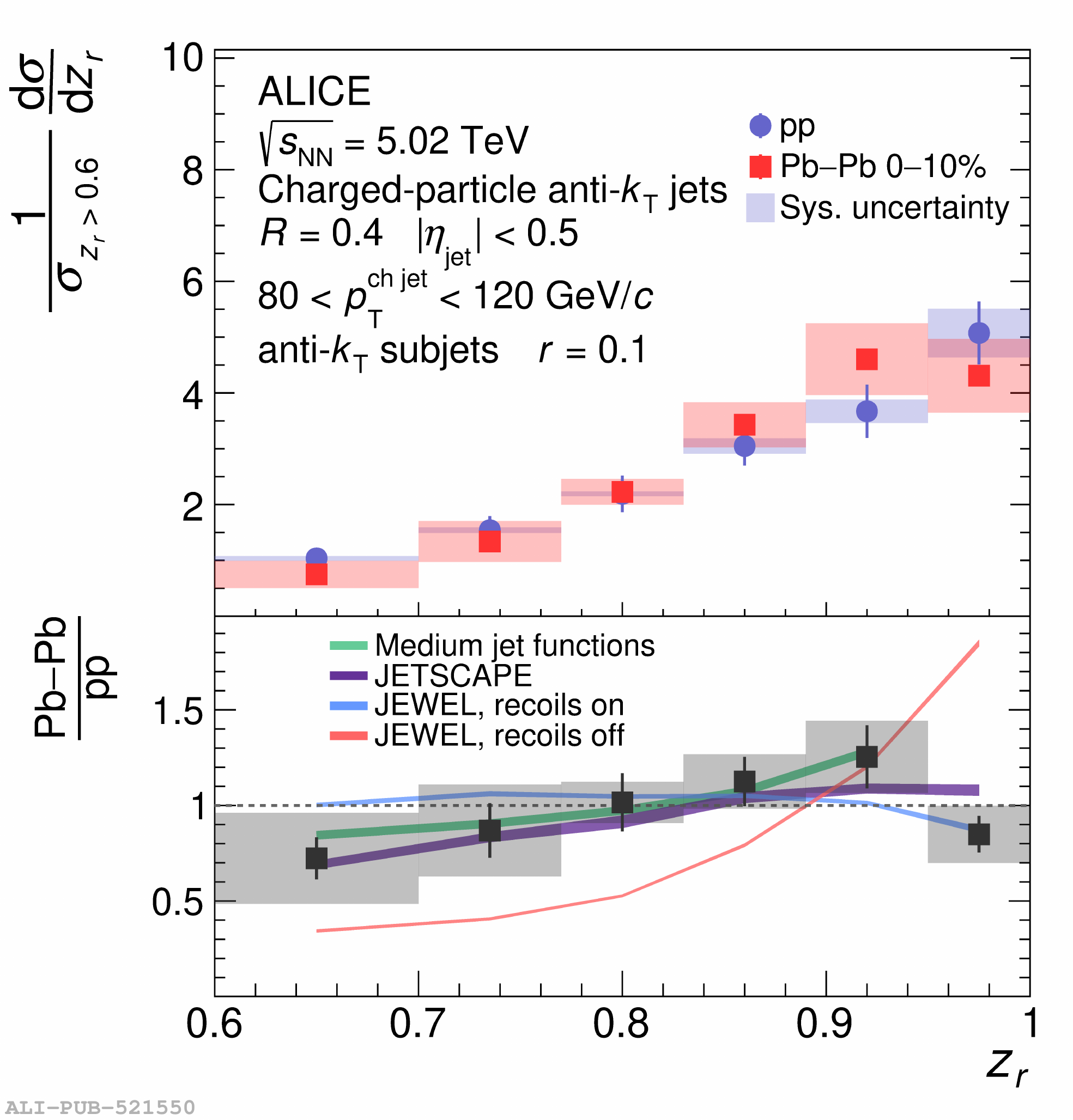}
    \includegraphics[width=0.48\textwidth]{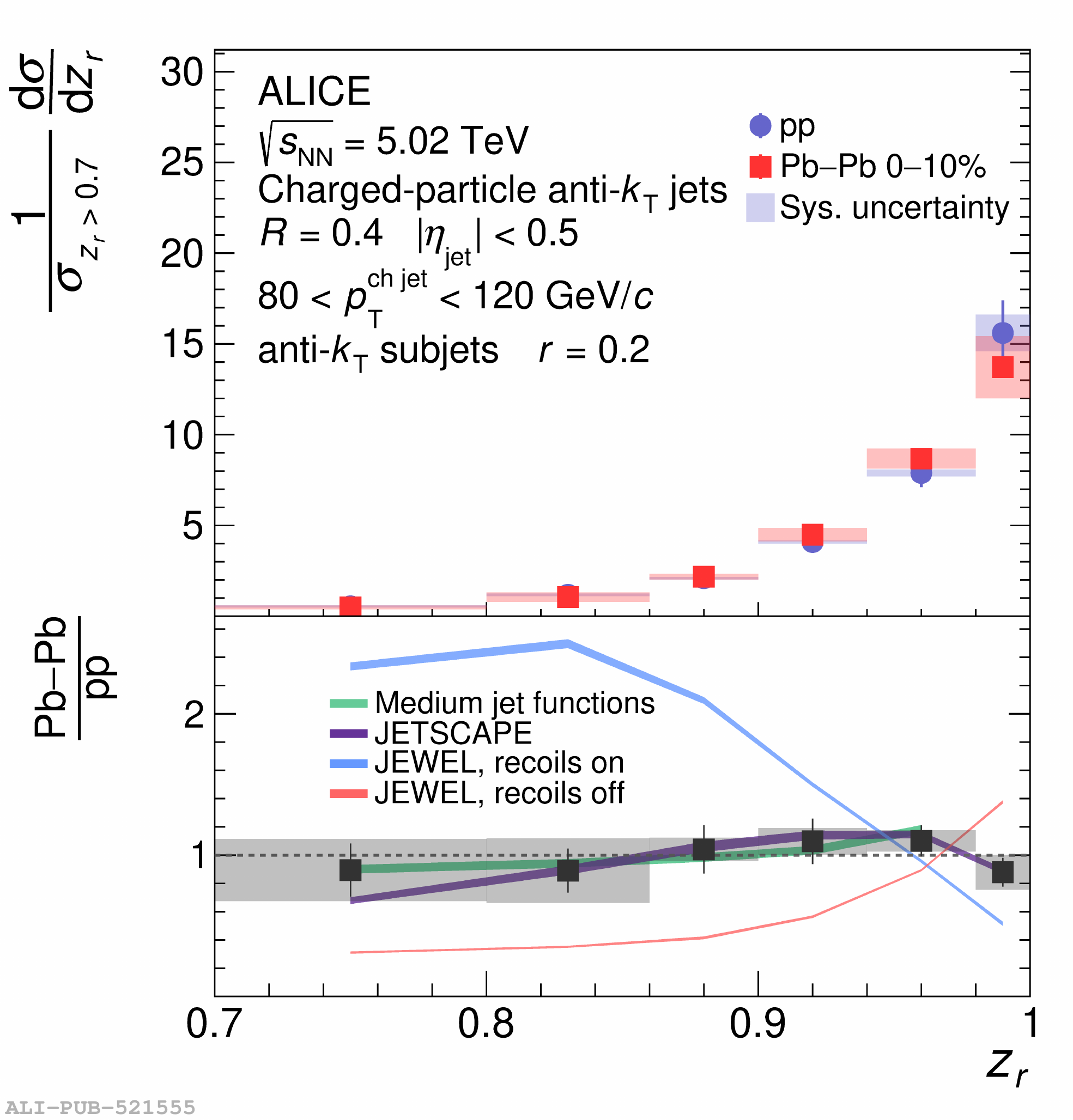}
    \caption{The ratio of subjet $z$ measured in $\pp{}$ and 0--10\% central $\PbPb{}$ collisions at \(\sqrts{}\) = 5.02 TeV for $R = 0.4$ charged-particle jets measured in $80 < \pTJetCh{} < 120$ \GeVc{}. The measurement for $r$ = 0.1 is shown on the left, while $r$ = 0.2 is shown on the right.}
    \label{fig:subjetZ}
\end{figure}

Ungroomed subjets can probe harder
fragmentation than is accessible solely via single hadrons. By
reclustering \(R = 0.4\) jets using the anti-\(\kT{}\) algorithm with
subjet resolution parameter \(r\), hard subjet properties can be
investigated. This study focuses on the subjet fragmentation,
\(z_{\text{r}} = \frac{p_{\text{T}}^{\text{ch,subjet}}}{\pTJetCh{}}\),
for the leading subjet identified in each jet \cite{ALICE:2022vsz}.
Figure \ref{fig:subjetZ} shows the ratio of subjet fragmentation in
0--10\% central \(\PbPb{}\) to that in \(\pp{}\) collisions for \(r\) = 0.1 (0.2) on the left (right). The data are consistent with no modification
in \(\PbPb{}\) collisions, although there is a hint of a change of shape
as \(z_{\text{r}} \to 1\) for \(r\) = 0.1. The interplay between the
softening at mid \(z_{\text{r}}\) due to medium-induced radiation and
the hardening at high \(z_{\text{r}}\) due to differences in the quark
and gluon fractions may lead to a similar non-trivial shape change. Both
the medium jet functions \cite{Kang:2017mda,Qiu:2019sfj} and JETSCAPE \cite{Putschke:2019yrg,Majumder:2013re,He:2015pra} curves describe the data fairly
well within their range of applicability, while JEWEL \cite{Zapp:2012ak,Zapp:2013vla} is not consistent
with the data except for \(r\) = 0.1 with recoils on.

\hypertarget{summary}{%
\section{Summary}\label{summary}}

ALICE has measured a wide variety of ungroomed and groomed jet
substructure observables in \(\PbPb{}\) and pp collisions at
\(\sqrts{} = 5.02\) TeV. Although there is no modification of the
groomed momentum splitting fraction \(\zg{}\), a narrowing effect is
seen for substructure observables with an angular dependence. The
underlying mechanism is still under investigation, with
possibilities including color coherence, changes in the quark and gluon
fractions in the medium, and bias towards narrower jets surviving
transversal of the medium. The distribution of groomed relative transverse momentum,
\(\kTg{}\), is the first application of Dynamical Grooming in heavy-ion collisions. However, it is not yet precise enough to be able to observe effects from Moliere scattering.

\scriptsize
\bibliographystyle{style/JHEP}
\bibliography{rehlers.ICHEP2022.bib}

\end{document}